\documentstyle [prl,aps,twocolumn]{revtex}
\newcommand{\la}[1]{\label{#1}}

\newcommand{\GG}{{\rm G}}

\newcommand{\n}{{\bf n}}

\newcommand{\be}{\begin{equation}}
\newcommand{\ee}{\end{equation}}
\newcommand{\ba}{\begin{eqnarray}}
\newcommand{\ea}{\end{eqnarray}}
\newcommand{\bastar}{\begin{eqnarray*}}
\newcommand{\eastar}{\end{eqnarray*}}
\newcommand{\half}{{1 \over 2}}
\newskip\humongous \humongous=0pt plus 1000pt minus 1000pt

\newif\ifdtup

\relax

\begin{document}
\title  {TOWARDS A STRING REPRESENTATION OF \\
         INFRARED SU(2) YANG-MILLS THEORY }
\bigskip

\author{Edwin Langmann$^{*}$ and 
Antti J. Niemi$^{** }$ } 
\address{$^*$Department of Theoretical Physics, Royal Institute of 
Technology, Stockholm, Sweden \\
$^{**}$Department of Theoretical Physics, Uppsala University, 
Box 803, S-75108, Uppsala, Sweden \\ 
$^{**}$Helsinki Institute of Physics, P.O. Box 9, FIN-00014 
University of Helsinki, Finland \\
$^{**}$ The Mittag-Leffler Institute, Aurav\"agen 17, S-182 62 
Djursholm, Sweden \\ 
{\scriptsize \bf LANGMANN@THEOPHYS.KTH.SE, NIEMI@TEORFYS.UU.SE } 
}

\maketitle

\begin{abstract}
We employ a heat kernel expansion to derive an 
effective action that describes four dimensional 
SU(2) Yang-Mills theory in the infrared limit. 
Our result supports the proposal that at large 
distances the theory is approximated by the
dynamics of knotted string-like fluxtubes which
appear as solitons in the effective theory. 
\end{abstract}

\narrowtext
\bigskip

Confinement remains an important unsolved problem in four dimensional 
Yang-Mills theory, even though several approaches have been developed 
for explaining it \cite{mitya1}. From the present point of view an 
interesting one is the construction of string variables that 
describe the theory at large distances \cite{sasha}. In \cite{fad1} 
an explicit string-like decomposition of four dimensional SU(2) 
gauge field has been introduced. Wilsonian renormalization 
group arguments suggest that in these variables the infrared 
SU(2) Yang-Mills theory is in the universality class of
\be
S(\n) \ = \ \int d^4x \ \left[  (\partial_\mu \n)^2 \ + \
\frac{1}{e^2} ( \n \cdot  \partial_\mu
\n \times \partial_\nu \n )^2 \ \right]
\la{eff}
\ee
where $\n$ is a three component unit vector. The action (\ref{eff}) supports 
string-like knotted solitons \cite{nature}, \cite{knots}, which 
suggests that at large distances SU(2) Yang-Mills theory describes
the dynamics of knotted fluxtubes. 
The first term in (\ref{eff}) is relevant in the infrared limit,
and it is uniquely defined. The second term
is marginal, and it stabilizes the solitons. Even though 
Lorentz invariance permits additional marginal contributions, the 
second term in (\ref{eff}) is special in the sense 
that time derivatives appear at most bilinearly. This ensures 
that (\ref{eff}) admits a Hamiltonian interpretation and as 
such it is unique.

In this Letter we present a first-principles heat-kernel
derivation of (\ref{eff}). We start from the SU(2) Yang-Mills 
Lagrangian
\be
L_{YM} \ = \ \frac{1}{4g^2} Tr F_{\mu\nu}^2
\la{YM}
\ee
where we decompose the gauge field as \cite{fad1}
\be
A^a_\mu \ = \ C_\mu n^a \ + \ \epsilon_{abc}
\partial_\mu n^b n^c \ + \ \rho  \partial_\mu n^a \ + \
\sigma \epsilon_{abc} \partial_\mu n^b n^c
\la{su2}
\ee
with $C_\mu$ a vector field and $\phi = \rho + i \sigma$
a complex scalar. Together they form an abelian Higgs multiplet,
covariant under U(1) gauge transformations in the direction 
$n^a$ of the Lie algebra SU(2). We show 
that if one integrates over this multiplet in the path integral, 
the ensuing effective action for the unit vector $\n$ 
is in the universality class of (\ref{eff}).  Indeed,
if we substitute (\ref{su2}) to (\ref{YM}), we get
\[
L_{YM} \ = \ \frac{1}{4g^2} ( \
[ G_{\mu\nu} - H_{\mu\nu} (1-|\phi|^2) ]^2 
\]
\be
+ [ ( v_{\mu\nu} \ + \ i H_{\mu\nu} )
(D_\mu\phi)^{*} D_\nu\phi + h.c \ ]
\ )
\la{YM1}
\ee
where we have defined the U(1) covariant derivative $
D_\mu \phi =  \partial_\mu \phi  +  i C_\mu \phi $,
and we have defined $ G_{\mu\nu} = \partial_\nu C_\mu - 
\partial_\mu C_\nu$, and $H_{\mu\nu} = \n \cdot \partial_\mu \n \times 
\partial_\nu \n$, and $ v_{\mu\nu} = \delta_{\mu\nu} 
(\partial_\rho \n )^2 - \partial_\mu \n \cdot \partial_\nu 
\n$. In particular, $H^2_{\mu\nu}$ is bilinear in $v_{\mu\nu}$,
\[
H^2_{\mu\nu} \ = \ \frac{1}{3} (v_{\mu\mu})^2 - v_{\mu\nu}
v_{\nu\mu}
\]
If we then average (\ref{YM1}) over $(C_\mu, \phi)$ with $ <|
\partial_\lambda \phi|^2  \delta_{\mu\nu}
- \partial_\mu \phi^* \partial_\nu \phi> \propto \ \delta_{\mu\nu}$ 
we get the Lagrangian in (\ref{eff}). Furthermore, if 
we average (\ref{YM1}) over $\n$ with $< H_{\mu\nu} > 
= 0$ and define $ < v_{\mu\nu} > =  m^2\delta_{\mu\nu}$ and
set $< H_{\mu\nu}^2 > = \lambda$ we get the abelian Higgs model 
\be
S \ = \ \frac{1}{4 g^2} \int d^4x \left\{  G_{\mu\nu}^2
+ m^2 |D_\mu \phi|^2 + \lambda (|\phi|^2 - 1)^2 \ \right\}
\la{abelian}
\ee
These Wilsonian arguments \cite{fad1} are suggestive, 
but leave a number of important issues unexplained. 
For example, (\ref{YM}) is (classically) scale 
invariant but (\ref{eff}) is not, and these arguments 
fail to explain how the mass scale 
appears. For a justification of (\ref{eff}) a 
first principles computation is needed. 
For this we redefine $C_\mu \to g C_\mu$ 
and $\phi\to g\phi$ and re-write (\ref{YM1}) as
\[
L_{YM} \ = \ \frac{1}{4} G_{\mu\nu}^2 \ + \
\frac{1}{4}(\frac{1}{g} -  g |\phi|^2)^2 H_{\mu\nu}^2 \
+ \ \frac{1}{2} v_{\mu\nu}\partial_\mu \phi^*
\partial_\nu \phi 
\]
\[ 
+ \frac{i}{2} g \ v_{\mu\nu}
[ (\partial_\mu \phi^*) \phi - \phi^* \partial_\mu \phi ]
C_\nu \
+ \ \half v_{\mu\nu} g^2 |\phi|^2 C_\mu C_\nu 
\]
\be
- C_\mu
\left( (\frac{1}{g} - g|\phi|^2  ) \partial_\nu H_{\mu\nu}  -3
H_{\mu\nu} g\partial_\nu |\phi|^2 \right)
\la{act1}
\ee
{}Here it is obvious that the combination $v_{\mu\nu}$
is the natural order parameter
in the effective action that follows when we integrate 
over $(C_\mu, \phi)$. Furthermore, since the ground state is both rotation 
and translation invariant, we conclude that the 
free energy {\it i.e.} effective potential can only depend 
on the constant part of the trace of $v_{\mu\nu}(x)$. 
This prompts us to redefine 
\be
v_{\mu\nu} ( x ) \ \to \ v_{\mu\nu} (x) \ + \ m^2 \delta_{\mu\nu}
\la{order}
\ee
where $m$ is a constant with the dimensions of a mass.
The effective action then admits a derivative expansion 
in powers of $v_{\mu\nu}(x)$ around $v_{\mu\nu} = m^2 \delta_{\mu\nu}$
and the leading term in this expansion, the effective potential, 
depends only on the rotation and translation invariant $m^2$.

We note that the second term in (\ref{eff}) already appears 
in (\ref{act1}) as the second term in its {\it r.h.s.}
In order to justify (\ref{eff}) it is then sufficient
to show that the first term in (\ref{eff}) is induced
when we integrate over $(C_\mu, \phi)$.
We also note that the second term in (\ref{act1})
is reminiscent of a Higgs mechanism, see also (\ref{abelian}).
This suggests the shift $\phi \to \phi  +  a$
with $a$ some {\it a priori} arbitrary complex parameter,
the {\it v.e.v.} of the complex scalar $\phi$.  
Consequently we introduce the parameter $ \Delta^2  =  
(\frac{1}{g} - g |a|^2)^2$ which measures the deviation 
of $|a|$ from $1/g$, the value of the complex scalar 
$|\phi|$ for which the second term on 
the {\it r.h.s.} of (\ref{act1}) vanishes.

We eliminate the U(1) gauge invariance from the
abelian Higgs multiplet by selecting the gauge fixing term
\be
L_{gf} \ = \ \frac{1}{2\xi} \left( \ M^{-2} \partial_\mu
(v_{\mu\nu}C_\nu) \ + \  \xi M^2
g \frac{i}{2} (\phi^* a - a^* \phi) \ \right)^2
\la{gf1}
\ee
This is a variant of the conventional $R_\xi$-gauge in
spontaneously broken gauge theories, with $M$ an arbitrary 
mass scale which we need to introduce since
in our units the field $\phi$ is dimensionless.
The corresponding ghost action is
\be
L_{ghost} \ = \ \bar\eta \ ( - M^{-2} \frac{1}{g}
\partial_\mu v_{\mu\nu} \partial_\nu \ + \ \xi M^2
g |a|^2 ) \  \eta
\la{ghost1}
\ee
The integration measure over $(C_\mu, \phi)$ is
determined as follows: We recall that geometrically 
${\cal A}_4$, the space of all four dimensional Yang-Mills 
connections $A^a_\mu$, is an infinite dimensional Euclidean
space with Cartesian flat metric which we write as
\[
ds^2 \ \sim \ {\cal G}(A,A) \ = \ \int d^4 xd^4 y \
\delta^{\mu\nu}_{ab}(x-y) dA_\mu^a(x) dA_\mu^b(x)
\]
Similarly, the {\it naive} path integral measure over
${\cal A}_4$ is an infinite dimensional analog of
the Cartesian measure $[dA] \sim \prod \ dA^a_{\mu}(x) $.
Geometrically the decomposition (\ref{su2}) defines an
embedding of a surface in ${\cal A}_4$, and integration 
over $(C_\mu, \phi)$ is with respect to the measure 
that is induced by this embedding.  
In order to determine this induced measure
we first define $ \n = \left( \cos \varphi  \sin \theta ,  
\sin \varphi  \sin \theta , \cos \theta \right)$.
The natural intrinsic measure for the components 
in (\ref{su2}) is then $[dC_\mu][\sin\theta d\theta][d
\varphi][d\rho] [d\sigma]$. But since (\ref{su2}) is 
also a parametrization of a surface which is embedded 
in ${\cal A}_4$, in addition we need to account for 
a  Jacobian $|\GG|^{\half}$ factor that emerges from the
embedding. This we determine as
follows: We first evaluate the variation of 
(\ref{su2}) {\it w.r.t.} its components
\[
dA^a_\mu(x) \ = \ \int d^4 y \{ \ \frac{\delta A^a_\mu (x)}
{\delta C_\nu(y)} dC_\nu(y) +  \frac{\delta A^a_\mu (x)}
{\delta \rho(y)} d\rho(y) 
\]
\[
+  \frac{\delta A^a_\mu (x)}
{\delta \sigma(y)} d \sigma(y) +  \frac{\delta A^a_\mu (x)}
{\delta \theta(y)} d \theta(y) +  \frac{\delta A^a_\mu (x)}
{\delta \varphi(y)} d\varphi(y)\ \}
\]
and then infer the Jacobian $|\GG|^\half$ factor by 
evaluating the determinant of the metric that we read from
the corresponding line element $ds^2$.  We interpret the result
as an additional contribution to our Lagrangian. In a 
derivative expansion in powers of $v_{\mu\nu}$ this gives 
us the contribution $ L_{measure}  =  - ln \left[ v^2_{
\mu\mu}(x) \right] + \ldots$
where we have ignored terms that do not contribute at the level
of the Gaussian approximation. The Lagrangian 
that we use in our computation of the effective action 
is then 
\[
L_{eff}  = L_{YM} + L_{gf}  + L_{ghost} \ + \ L_{measure}
\]
We compute the path integral over the fields $(C_\mu, \phi, \eta, 
\bar\eta)$ using standard heat kernel expansion. 
Details of the computation are straightforward but somewhat 
tedious, and we have used Maple to perform symbolic manipulations.
After combining the various functional determinants that emerge
from the Gaussian integrations we conclude that to leading order 
in a gradient expansion in $v_{\mu\nu}$ the effective action is
\[
L_{eff} \ = \ \ \frac{\Delta^2}{4}
H^2_{\mu\nu}(x)
\]
\be 
+ \ \int\limits_{|p|\leq \Lambda}
\frac{ d^4 p}{(2\pi)^4} \left( \ \half
ln det[ {\hat Q}^C_{\mu\nu}] \ + \ \half ln [{\hat Q}] \ - \
ln [v_{\mu\mu}] \ \right)
\la{eff1}
\ee
where $ {\hat Q}^C_{\mu\nu}(x) = p^2 \delta_{\mu\nu}  -
p_\mu p_\nu + g^2 |a|^2 v_{\mu\nu}(x)$ and $
{\hat Q} = v_{\mu\nu}(x) p_\mu p_\nu + 2 g^2
H^2_{\mu\nu}$ and where we also need to perform 
the shift (\ref{order}).
Here $\Lambda$ is an ultraviolet cut-off that
we introduce in order to regulate the 
momentum integral. We note that the result
is {\it manifestly} U(1) gauge invariant {\it i.e.} it is
independent of the parameters $M^2$ and $\xi$ in
(\ref{ghost1}), as it should.

The evaluation of the momentum integral in (\ref{eff1})
is straightforward. We describe the final result as
\[ 
L_{eff} = \frac{\Delta^2}{4} H^2_{\mu\nu}(x) + 
\left\{ \ L^{(0)}(m) + L^{(2)}(v) +
L^{(4)}(v) \right\} 
\]
where $L^{(0)}(m)$, 
$L^{(2)}(v)$ and $L^{(4)}(v)$ are of
zeroth, first and second order in powers of $v_{\mu\nu}$,
in a derivative expansion.
These contributions can all be presented using the function
\[
f(u) \ = \ \frac{1}{8} + \ \int_0^1 d\xi \ \xi^3 \ ln (\xi ^2 + u)
\]
\[ 
= \ \frac{1}{4} u \left[ 1 - u \cdot ln( \frac{1+u}{u}) \ 
\right]
\ + \ \frac{1}{4} ln ( 1 + u)
\]
and its derivatives, where $u  =  (\frac{m}{\Lambda})^2 $ is a
dimensionless parameter. We note that the function $f(u)$ is 
non-negative and monotonically increasing, 
with $f(u\to 0) = {\cal O}(u)$ and $f(u \to \infty) = {\cal 
O}\left( ln (u)\right)$. 

For the zeroth order contribution $L^{(0)}(m)$
we have the following $u$ dependence,
\be
L^{(0)} \ = \ \frac{\Lambda^4}{16 \pi^2} \left\{ \frac{16 \pi^2}{3}
\Delta^2 u^2 \ + \ 3 f ( g^2 |a|^2 u )
\ + \ f ( \frac{8}{3} g^2 u) \  \right\}
\la{S0}
\ee
This is the vacuum (Casimir) energy in the Gaussian approximation.
It has the expected ${\cal O}(\Lambda^4)$ divergence
multiplying the (uniform) vacuum energy density. The ground state
values of the various parameters in the background of
a constant $\n$ can be determined by minimizing (\ref{S0}).
This suggests that we set $u = 0$ so that (\ref{S0}) vanishes. 
However, we note that for a fully reliable minimization one needs to 
account for certain additional terms. In particular one 
needs to include the full ghost contribution 
that comes from fixing the full SU(2) gauge invariance. 
For this we need to understand how (\ref{su2}) emerges from 
the conventional gauge fixing procedure, which is beyond 
the scope of the present letter.

The first-order contribution $L^{(2)}(v)$ yields in the leading
order our desired first term in (\ref{eff}),
\[
L^{(2)}(u) \ = \ \frac{ \Lambda^2}{64 \pi^2} \{
\frac{32}{3} \pi^2 \Delta^2 u \ + \ 3 g^2|a|^2 f' ( g^2|a|^2 u) 
\] 
\be
+ \ \frac{8}{3} g^2 f' ( \frac{8}{3} g^2 u) \}
\cdot (\partial_\mu \n)^2
\la{S2}
\ee
The derivative of $f(u)$ is monotonically
decreasing, with $f'(0) = \half$ and $f'(u \to \infty) \to 0$.
In particular, when $u=0$ and  
the Casimir energy (\ref{S0}) vanishes, 
the coefficient in (\ref{S2}) remains non-zero.
Indeed, since $f'(u)$ is non-vanishing 
and positive, the 
coefficient in (\ref{S2}) is always 
positive and its minimum
value occurs for a {\it non-vanishing} value of $u$. 
This suggests that in order to locate the ground state 
in the background of a nontrivial $\n$ we should
select $u \not =0$. This means that $m \not=0$ and  
it scales nontrivially in $\Lambda$. 
But for our present purposes the exact 
determination of the ground state value of $u$, or the
scaling of $m$ in $\Lambda$ is inessential. Here it is 
sufficient to note that for all $u$ 
(\ref{S2}) is positive and non-vanishing. This 
means we have a mass gap and the first term in (\ref{eff}) 
is present, irrespectively of the exact value of $u$. 

Since (\ref{S2}) is infrared relevant and the second, 
infrared marginal term in (\ref{eff}) appears 
in (\ref{act1}), we may already at this point 
conclude that (\ref{eff}) is indeed an effective action 
that approximates the infrared limit of the SU(2) 
Yang-Mills theory. But for completeness we 
also evaluate the Gaussian correction to the
coefficient of the second term in (\ref{eff}).
At this point it becomes apparent that there is 
certain latitude in the definition of the coefficient 
multiplying $H_{\mu\nu}^2$: The natural 
order parameter for the effective
action is $v_{\mu\nu}$, and $H_{\mu\nu}$ is but a particular
bilinear combination of this order parameter. 
There are also various additional bilinear
combinations of $v_{\mu\nu}$ that can be present, and here we select
a particular basis of such bilinears 
that allows us to write the leading contribution as
\be
L^{(4)}(u) \ = \ \frac{1}{16 \pi^2} ( 4 \pi^2 
\Delta^2 \ - \ A) H_{\mu\nu}^2
\ + \ \frac{1}{16 \pi^2} (\frac{A}{3}-B)
v_{\mu\mu}^2
\la{S4}
\ee
where
\[
A \ = \ \frac{1}{192 u^2} \ + \ \frac{4}{9} \frac{g^2}{u}
f'( \frac{8}{3} g^2 u)  \ + \ \frac{4}{27} g^4 f '' (\frac{8}{3} g^2 u)
\]
\[ 
+ \ \frac{g^2 |a|^2}{24 u} f'(g^2 |a|^2 u) \ + \
\frac{1}{48} g^4 |a|^4 f''(g^2 |a|^2 u)
\]
and
\[
B \ = \  \frac{1}{48 u^2} \ + \ \frac{16}{9} \frac{g^2}{u}
f'( \frac{8}{3} g^2 u)  \ - \ \frac{8}{27} g^4 f '' (\frac{8}{3} g^2 u)
\]
\[ 
+ \ \frac{g^2|a|^2}{6 u} f'(g^2|a|^2 u) \ - \
\frac{7}{24} g^4 |a|^4 f''(g^2|a|^2 u)
\]
For the second derivative we have $f''(0) = - \infty$ and
$f''(u \to \infty) \to 0$.  Depending on the relative value of
parameters, the coefficient of $H_{\mu\nu}^2$ is typically
negative when $u < u_0 < 1$ and positive when $u_0 < u < 1$ for
some $u_0 >0$, while the coefficient of the second term 
in (\ref{S4}) is negative. Since the energy density should 
be positive this yields 
restrictions on the possible values of the 
parameters, and suggests that we should set $u > u_0$. 
We note that this resembles the situation 
in \cite{mitya2}. In analogy we then conclude that
the negative value of the energy density (\ref{S4}) for 
some range of parameters is an indication that higher order terms
in the derivative expansion become important. Indeed, one can show that the 
{\it full} energy density is necessarily positive 
for all values of the parameters.

The two terms that are present in (\ref{S4}) are both 
local and Lorentz-invariant. Both are also marginal in
the infrared, but since the second term in (\ref{S4}) has
a fourth order contribution from time derivatives 
it does not admit a canonical Hamiltonian interpretation. 
Consequently (\ref{eff}) is unique in the sense that it 
admits a Lorentz invariant Hamiltonian interpretation.
In particular we may view (\ref{eff}) as a Hamiltonian that
represents the universality class of infrared SU(2) 
Yang-Mills theory. But we have
also found that there is some latitude in the definition
of the coefficient multiplying $H_{\mu\nu}^2$. In modelling 
the dynamical details of large distance SU(2) Yang-Mills 
theory, it is then necessary to include additional 
infrared marginal terms. We note that this is parallel
to the situation in the Skyrme model, where one should
account for all three terms with four derivatives. 
  
Finally, we verify that (\ref{eff}) together with its
knotted string-like solitons is indeed consistent with 
certain familiar aspects of non-perturbative Yang-Mills 
theory. For this we remind that in the
temporal $A_0=0$ gauge Yang-Mills instantons
interpolate between asymptotic $t=\pm \infty$ vacua,
classified by  $\pi_3(S^3) \sim Z$ homotopy classes. 
The difference in the $t=\pm \infty$ homotopy classes 
coincides with the second Chern character of the 
interpolating instanton configuration.
For consistency, this picture should be 
reflected in the topological properties 
of the knotted solitons: 
If we substitute the decomposition (\ref{su2}) 
in the three dimensional Chern-Simons 
action that counts the $t = \pm \infty$
homotopy classes, we find that for a flat $A_0 = 0$ connection
the Chern-Simons action yields the Hopf invariant of the 
configuration $\n$,
\[
\omega_3(A) \ = \ \epsilon_{ijk}Tr\{ A_i \partial_j A_k +
\frac{2}{3}A_i A_j A_k \} \ = \ \frac{1}{4} \epsilon_{ijk}
C_i H_{jk}
\]
with $ H_{ij}  =  \partial_j C_j - \partial_j C_i  =
\n \cdot \partial_i \n \times 
\partial_j \n$. 
Consequently there is a relationship between the $A_0=0$ instanton
structure of Yang-Mills theory and the (self-)linking number of
knotted solitons in the effective infrared theory. In particular, 
instantons can be viewed as configurations that interpolate 
between different knotted vacuum configurations in $\n$.  
We suggest that this relationship is additional evidence 
supporting the consistency of the
present interpretation of the infrared structure 
of SU(2) Yang-Mills theory. 

\vskip 0.4cm
In conclusion, we have studied the large distance structure
of four dimensional SU(2) Yang-Mills theory. By employing a 
heat-kernel expansion of the effective action we have shown 
that the Lagrangian (\ref{eff}) indeed follows from the
decomposition (\ref{su2}) of the gauge field. Our result 
supports the proposal made in \cite{fad1}, that at large
distances SU(2) Yang-Mills theory becomes an effective 
string theory, where the strings appear as solitons
that are stabilized against shrinkage by their nontrivial,
knotted topology. We have also verified that this 
interpretation is consistent with the familiar 
instanton $\theta$-vacuum structure of SU(2) Yang-Mills theory.

\vskip 0.4cm

{\bf Acknowledgements} A.N. thanks Ludvig Faddeev for numerous 
discussions, and  D. Diakonov for valuable suggestions.
The research by A.N. was partially supported 
by NFR Grant F-AA/FU 06821-308.

\end{document}